# Software Architecture Decision-Making Practices and Challenges: An Industrial Case Study

Sandun Dasanayake, Jouni Markkula, Sanja Aaramaa, Markku Oivo
M-Group, Faculty of Information Technology and Electrical Engineering, University of Oulu
Oulu, Finland
{sandun.dasanayake, jouni.markkula, sanja.aaramaa, markku.oivo}@oulu.fi

*Abstract*—Software architecture decision-making is critical to the success of a software system as software architecture sets the structure of the system, determines its qualities, and has far-reaching consequences throughout the system life cycle. The complex nature of the software development context and the importance of the problem has led the research community to develop several techniques, tools, and processes to assist software architects in making better decisions. Despite these effort, the adoption of such systematic approaches appears to be quite limited in practice. In addition, the practitioners are also facing new challenges as different software development methods suggest different approaches for architecture design. In this paper, we study the current software architecture decision-making practices in the industry using a case study conducted among professional software architects in three different companies in Europe. As a result, we identified different software architecture decision-making practices followed by the software teams as well as their reasons for following them, the challenges associated with them, and the possible improvements from the software architects' point of view. Based on that, we recognized that improving software architecture knowledge management can address most of the identified challenges and would result in better software architecture decision-making.

## I. INTRODUCTION

Decision-making is an integral part of software development, and various decisions are being made throughout the software development life cycle concerning processes, products, tools, methods, and techniques [1]. However, software architecture decisions carry additional weight compared to other decisions since many of the architecture decisions are made in the early stage of the software development life cycle and have significant influence on shaping the analysis of the problem and the expression of the design [2]. Despite being fundamental to the system and hard to change, architecture decisions also facilitate managing and reasoning about the changes as the system evolves [3]. Even though design decisions used to be implicitly embedded in the architecture, describing architecture as a set of design decisions is gaining recognition as one widely accepted definition [4] [5].

A considerable number of software architecture decision-making techniques have been developed from different perspectives in order to make software architecture decisions in a systematic way [6] [7]. In addition, a range of research work has also been carried out on improving architecture knowledge management, collaboration, and documentation to provide support during architectural activities [8] [9] [10]. However, software architects have found it difficult to make architecture decisions for reasons such as dependencies on other decisions, the major business impact caused by the decision, serious negative consequences resulting from the decision, and the amount of effort required to analyze the possible alternatives [11]. Recent studies carried out among professional software architects suggest that software architects tend to use their own customized decision-making approaches rather than using systematic architecture decision-making approaches from the literature [12] [13].

Practitioners are facing new challenges related to architecture decision-making as software development is undergoing rapid changes with the increased use of development practices such as agile software development, DevOps, and continuous delivery. While traditional software development approaches emphasize up-front architecture design, modern software development methods favor continuous design where the architecture evolves as the system development progresses [14]. Finding the right balance between the up-front design and its evolution during the software development life cycle is vital for the success of system development [15] [16].

In this paper, we present a case study carried out to investigate how software architecture decisions are made by professional architects in an industrial context and analyze the results in order to identify the different architecture decision-making approaches followed, the reasoning behind using them, the challenges associated with them, and the possible improvements that can be made to achieve better architecture decision-making.

After the introduction, Section II describes the related work with the theoretical aspects of decision-making and knowledge management in general as well as in the context of software architecture. Then Section III describes the research method, including the various steps of the case study process. While Section IV showcases the results of the study and analyzes them in light of the research questions, Section V interprets the results using the research literature. Finally, Section VI presents possible validity threats, and Section VII concludes the paper by highlighting the important elements of the study and future work.

## II. RELATED WORK

Decision-making is one of the basic cognitive processes of human behavior by which a preferred option or a course of action is chosen from among a set of alternatives based on certain criteria [17]. It has been studied from different aspects in different disciplines; hence, there are many different taxonomies available based on the different aspects of decision-making. There are two main paradigms of decision theory, normative and descriptive. While the goal of normative decision theory is to identify the best possible decisions assuming that a well-informed rational decision-maker would adhere to a well-defined process, descriptive decision theory attempts to uncover the strategies and cognitive process underlying how decisions are actually made in real-life scenarios [18]. The actual decision-making process can significantly deviate from the best possible process because heuristics and biases affect human decision-making [19]. Naturalistic decision-making is a descriptive approach to studying how humans make decisions in complex real-world situations with several constraints [20].

Since decision-making is a knowledge-intensive activity with knowledge as its raw materials, work-in-process, by-products, and finished goods [21], decision-makers who make crucial decisions should possess knowledge about the decision problem as in other relevant areas. Knowledge can be identified in two different forms as explicit knowledge or tacit knowledge, depending on its characteristics. Explicit knowledge refers to knowledge that can be transmitted in formal, systematic language, while tacit knowledge has a personal quality that makes it difficult to formalize and communicate [22]. Explicit knowledge can be found in wikis, textbooks, manuals, and other forms of audiovisual media. On the other hand, tacit knowledge is more personal and is frequently referred to using epitomes such as intuition, skills, insight, know-how, beliefs, mental models, or practical intelligence [23]. Even though the tacit-explicit classification is widely accepted, there are different views about the distinctions between tacit and explicit knowledge. While some scholars consider knowledge as having two distinct categories, others consider it a continuum where tacit and explicit knowledge represent the two extremes of the spectrum [24] [25]. While implicit knowledge can be placed between tacit and explicit knowledge [26], it is also possible to integrate it into a continuum based on the degree of codifiability [27]. On the other hand, having entirely explicit knowledge is not feasible since explicit knowledge should be tacitly understood and applied to be useful [28].

Knowledge in organizations is not limited to documents or repositories, it is also embedded in organizational routines, processes, practices, and norms [29]; hence, managing knowledge requires a holistic approach rather than merely collecting and sharing knowledge from various sources. Even though it is possible to follow a preemptive process to manage explicit knowledge, managing tacit knowledge requires actions dependent on the context. In an organization, explicit knowledge signifies the "process" that is concerned with how knowledge is organized, whereas tacit knowledge represents "practice," which refers to how work is done [30]. Several theoretical models conceptualize knowledge management activities from different perspectives. The SECI model, the Sensemaking model, the Wiig model, and the I-Space model are some of the widely used knowledge management models [31]. Among them the SECI model assumes that knowledge is created by interaction of tacit and explicit knowledge and presents four modes of knowledge conversion: socialization (from tacit knowledge to tacit knowledge), externalization (from tacit knowledge to explicit knowledge), combination (from explicit knowledge to explicit knowledge), and internalization (from explicit knowledge to tacit knowledge) [32].

### A. Architecture Decision-Making

The primary architecture decision-making task is making design decisions during the architecture design process. However, the interactions between software architecture and architecture decisions are not limited to the design phase. Architecture decisions play a crucial role in software architecture during various stages of the software system's life cycle, including development, evolution, and integration [33]. Hofmeister et al. [34] present three architecting activities where major architecture decision-making takes place: architectural analysis, architectural synthesis, and architectural evaluation. Architectural analysis serves to define the problem that is to be solved by the architecture; hence, decisions should be made regarding selecting requirements, prioritizing them, and analyzing them. The possible solution to the defined problem is proposed during architectural synthesis. This is the core of the architectural activities where the major decisions related to the architecture should be made. During architectural analysis, the proposed solution is analyzed in terms of the requirements, and decisions are made regarding its readiness and further improvements.

Software architecture decision-making is an inherently complex task since the architecture should address various stakeholder concerns in order to achieve system development goals. The quality attributes that should be fulfilled by the software system and the interaction between them are among the main factors that should be taken into consideration during architecture decision-making because architectures allow or preclude almost all the quality attributes of the system [35]. Conflicting and crosscutting concerns that are commonly seen in modern-day systems add further complexity to decision-making [36] [37]. Software architecture decision-making is primarily considered to be the software architect's responsibility [38]. Nevertheless, the active involvement of other stakeholders during the decision-making process is crucial to having a better understanding of the criteria that should be fulfilled by the architecture. A number of techniques, tools and processes have been proposed to assist software architects as well as software teams in making architecture decisions. ATAM [39], CBAM [40], the Quality-Driven Decision Support Method [41], and ATRIUM [42] are some of the well-known techniques.

### B. Architecture Knowledge

Software architecture knowledge consists of the architecture design itself together with the design rationale, design decisions, assumptions, context, and other factors that determine the nature of the architecture [43]. In addition to that, other abstract knowledge sources such as architecture styles and patterns, design patterns, and architecture and design tactics can be applied in different contexts [44]. However, for the most part, architecture knowledge resides inside the decision-maker's head as tacit knowledge. Since the main factors that drive software architecture design—reusing existing solutions, following a systematic method, and making decisions based on the decision-maker's intuition [7]—are knowledge-intensive activities, architecture knowledge plays an important role during architecting. While using a systematic method primarily requires implicit knowledge, intuition-based decision-making is largely based on tacit knowledge. On the other hand, reusing an existing solution requires both explicit and tacit knowledge based on the context.

Architecture knowledge management that identifies and captures all forms of architectural knowledge and makes it available for transfer and reuse across projects in an organization helps improve the outcome of the architecture process [45]. Documenting software architecture knowledge into design documentation is one of the most widely used architecture knowledge management activities [46]. However, just documenting the architecture without the design rationale and the contextual information will considerably undermine its usefulness. There are also many other types of architecture knowledge management activities, such as using knowledge repositories, wikis, knowledge management tools, forums, social media, formal and informal meetings, brainstorming and retrospective sessions, and collaborative working. Despite the importance of architecture knowledge management, there are many obstacles to making it work in practice due to various reasons—lack of motivation on the part of the stakeholders when the benefits do not seem to justify the effort, short-term project interests outweighing long term gain from architecture knowledge management, some decisions being made without reflection, and organizational structures and practices that hinder knowledge sharing [47].

## III. RESEARCH METHOD

The context the decision problem is embedded in plays an important role during decision-making [48]. Hence, only studying the decision process is not sufficient to gain a holistic understanding of decision-making. In the case of software architecture, many contextual factors influence decision-making including, but not limited to, stakeholder interactions, organizational culture, and resource constraints. Thus, it is important to study software architecture decision-making in its real-life context. Since a case study allows us to get insights about a given phenomenon as well as the related context, a case study was selected as the research method employed in this study.

### A. Case Study Design

Having a proper case study design that covers the various elements that should be considered during the study is crucial for a successful case study [49]. First, a case study design that includes the objective of the study, case description, theoretical framework, research questions, data collection and analysis strategies, and validity aspects was laid out. As expected during an empirical study, some elements of the case study design were reassessed and updated as the study progressed.

TABLE I. CASE COMPANY INFORMATION

| Company | Employees | Business Area | Team Size |
|---|---|---|---|
| A | ≈ 60 | Solutions for developing and transforming software systems | 1 - 5 |
| B | ≈ 70 | Research and development for aerospace and security domains | 1 - 5 |
| C | ≈ 50 | Tools and services for systems modeling, analysis, and validation | 1 - 7 |

However, the case study design helped to anticipate and accommodate changes while maintaining the rigor and the focus of the study. While the major part of the case study design is described here, some of the elements such as theoretical framework and validity aspects are discussed in the relevant sections elsewhere in the paper.

This study uses a multiple-case design as the evidence from the multiple cases is often consider more compelling, and therefore improves the robustness of the overall study [50]. The case study took place in three software development companies in two European countries during September and October 2014. As shown in Table I, all three companies are in the same category based on the number of employees, and the project team size is also in the same range.

The objective of this case study is to understand the current state of software architecture decision-making in the industrial context in order to provide decision support for making better architecture decisions. The following research questions are defined based on the given objective.

- RQ1: How do the software architects make architecture decisions?

- RQ2: What are the reasons for using the current architecture decision-making approach?

- RQ3: What challenges are associated with the current architecture decision-making approach?

- RQ4: Which areas can be improved in order to make better architecture decisions?

The aim of the first research questions (RQ1) is to have a holistic view of software architecture decision-making in the given context. The plan is to derive information about the various aspects of architecture decision-making including the process, techniques, and factors that decision-making is based on. The second question (RQ2) is designed to find out why the current architecture decision-making approach is used by the practitioners. While the third question (RQ3) addresses the different challenges related to the current decision-making approach, the fourth question (RQ4) targets areas that need to be improved during future research.

### B. Data Collection

Two data sources were used to obtain the data for the case study: the semi-structured interviews and the documents that are used or created during architecture design. While this approach helped increase the amount of data, it also served to increase the precision by data source triangulation [51].

TABLE II. INTERVIEWEE INFORMATION

| ID | Position | Role / Responsibility (in addition to architecting) | Experience |
|---|---|---|---|
| A1 | Senior Consultant | Project management, Software development | 16 years |
| A2 | Team Manager | Team management, Software development | 15 years |
| A3 | Software Engineer | Software development | 5 years |
| A4 | Software Engineer | Software development | 2 years |
| B1 | Software Engineer | Project management, Software development | 8 years |
| B2 | Software Engineer | Team management, Software development | 15 years |
| B3 | Software Engineer | Project management, Software development | 6 years |
| B4 | Software Engineer | Project management, Software development | 6 years |
| C1 | Software Product Lead | Product / Project management | 24 years |
| C2 | Software Engineer | Project management, Software development | 8 years |

The main criterion for selecting the interviewees was that they should have the responsibility for making software architecture decisions. Table II shows the position, responsibility, and experience of each interviewee. Even though none of the companies employ designated architects, we use the term *architects* to refer the interviewees since all of them assume the roles and responsibilities of software architects [52] and make architectural decisions in their current teams. All the participants belong to different project teams in their respective companies.

An interview guide with primary interview topics and guiding questions was prepared to help the researchers to carry out the interviews. As the name suggests, it was used as a guidance tool rather than a fixed questionnaire. The questions were designed to be open ended, and more detailed questions were improvised during the interviews, meaning subsequent questions or comments were formed based on words and phrases used by interviewees to reflect their opinions. As shown in Table III, the interview guide consisted of several sections with each section targeting a different type of information relevant to the study. Prior to conducting the interviews, the guide was evaluated and improved by two other senior researchers. It was also validated by conducting a pilot interview with a software engineering researcher who has many years of industrial experience as a software architect.

The interviews were carried out as face-to-face interviews on site, and each interview lasted from one to two hours. Prior to each interview, the researchers briefly explained the objective of the case study as well as the other issues such as the data handling, and privacy. An open discussion was held at the end of each interview to discus the highlights and make sure that there were no misinterpretations. All the interviews were recorded with the consent of the interviewees and later transcribed for analysis. The companies were requested to provide the documents relevant to software architecture prior to the interviews, and some documents mentioned during the interviews were also obtained later.

TABLE III. INTERVIEW GUIDE SUMMARY

| Section | Targeted Information |
|---|---|
| Case context | Practitioner's title, experience, role<br>Project goals, size, nature<br>Company background, business area, teams<br>Software development life cycle |
| Architecture design | Architecture design phase in brief<br>People involved and their roles<br>Architecture evaluations evolution and reuse |
| Architecture decision-making | Decision-making process<br>People involved and their roles<br>Documents used and produced<br>Design rationale<br>Manage/share/reuse information and knowledge<br>Advantages, disadvantages, and improvements |
| Decision-making techniques / tools | Currently used techniques and tools<br>Awareness of techniques and tools<br>Reasons for using or not using them<br>Advantages, disadvantages, and improvements |
| Expectations | Identified improvement areas<br>Expected improvements<br>Characteristics of the expected solution |

*C. Data Analysis*

The interview transcriptions, relevant documents collected from the companies, and related research literature were imported into the NVivo qualitative analysis tool. The analysis started with the themes created based on the research questions, and the researchers went through each document and labeled different segments of the text with the code of each theme. Since the themes are not mutually exclusive, some information was categorized into multiple themes. Though it appears to be redundant, having all the information related to a given theme is necessary for having a good understanding of the subject. As the coding progressed, new information emerged from the data and was categorized into new themes.

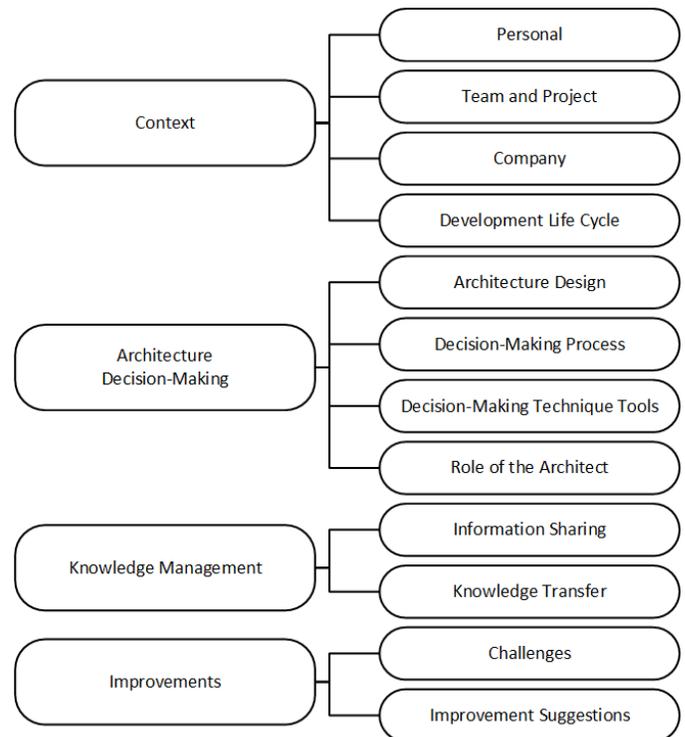

Fig. 1. Themes and subthemes

Fig. 1 shows the final set of themes and subthemes created during this study. Knowledge management is one important theme that emerged during the data analysis. Initially, all the possible improvements were categorized under the "Improvements" theme. However, it was evident that this theme was dominated by the information related to knowledge management, so it was separated as a different theme and recognized as a major improvement area identified during the study. Later the information under each theme was synthesized to form a detailed understanding of each area. If some information was unclear or the context was missing, the transcribed interview and the audio recording were used to clarify them. Throughout the case study, special attention was paid to protecting the companies' and individuals' privacy and integrity.

## IV. RESULTS

Based on the goals of the study, the interviews and the collected documents were analyzed focusing on case context, architecture decision-making, and knowledge management. The results are arranged according to the research questions. Section A describes the general case context while Sections B through E address RQ1 to RQ4 respectively.

### A. Case Context

It is important to understand the organizational structure, the nature of the projects, and the software development process before investigating the software architecture decision-making. All three case companies can be categorized as small to medium-sized enterprises (SME) based on the number of employees and turnover. According to the interviewees, all of them have an organizational structure where software development teams are free to make most of the decisions related to their activities themselves. Each company is involved in two different types of software projects, developing software products and providing software services to clients. Generally, customer involvement in the product development activities is minimal, but the service projects constantly interact with customers.

Even though a company-wide software development process was not followed in any of the case companies, the practitioners claim to use either a traditional or an agile-like development process. The agile-like development process is preferred among the practitioners; however, some of them are forced to use the traditional approach due to customers' preferences. In the case of agile-like development, they use selected elements of agile software development, such as sprint-based development. The requirement elicitation, analysis, and management are done by the software development team together with either an internal or external customer depending on the project type. There is also a strong emphasis on unit testing as a mechanism to ensure quality, especially in the case of customer-facing projects.

### B. Architecture Decision-making (RQ1)

Similar to the software development process, none of the case companies has a company-wide process or guidelines on software architecture design. However, regardless of the company and the size or type of the project, the importance of software architecture was widely recognized among the practitioners. However, two types of architecture design approaches followed by the practitioners were identified.

- Up-front design approach: The larger and waterfall-like projects tend to follow this approach. The team spent considerable time on the design phase and laid down a detailed architecture design. The design can still be changed as the project progresses, but the changes are minor since the team has already covered most of the design aspects.

- Continuous design approach: The smaller and agile-like projects follow this approach. They start with a minimal design that is expanded as the project progresses. The team spends less time deliberating different aspects of the initial architecture design.

While the upfront design approach provides clear guidance to the developers, its lack of flexibility can slow down the progress. If there is a flaw in the initial design, there is a risk of scrapping the whole project and start over in the middle of the development. On the other hand, the continuous design approach is flexible and ready to accommodate changes as the project progress. Even though missing initial design can run into the loss of focus, it provides benefits such as simple design, possibility to add new features and little duplication. Most of the projects described by the practitioners, follow a hybrid of the above two approaches. They have considerable design up front, however they continue to modify it based on their learning throughout the project.

Whether it is the upfront design approach or the continuous design approach, the practitioners have to make certain choices while creating a design. Even though every team has someone responsible for architecture design decisions, none of them has a traditional software architect whose sole duty is creating and maintaining architecture design. Most of the time the architect's role is assigned to the most experienced person on the team unless there is a compelling reason. According to the interviewees, all the teams get their members involved during the design decision-making process. While most of the project teams prefer to follow a consensus decision-making approach where all the members in the team give their consent to the selected choice, three of the practitioners mentioned that the architect takes the final decision in their respective teams. The architect's responsibility to ensure the system quality was cited as the main argument in favor of that approach. The architect's expertise in the relevant area was also another reason. When it comes to the formality of the decision-making process, only one out of ten software architects claims to have used at least a semi-formal method, while all the others claim that their approaches are informal.

The decision-making techniques used by the various teams can be categorized into three different groups: selecting a choice that fulfills pre-defined criteria, selecting a choice by analyzing pros and cons, and selecting the first satisfactory choice. Since these are not mutually exclusive, it is possible to combine approaches in selecting a solution.

- A choice that fulfills pre-defined criteria: This was the most common approach since five of the ten

interviewees (A2, A3, A4, B2, C2) claimed their teams use this approach to make design decisions. However the level of the criteria definition varies among the teams; one team uses extensibility as the only criterion while the others use multiple criteria based on the context. This is a rational approach, but it is restricted since a limited number of criteria are used to make the decision.

- A choice selected by analyzing pros and cons: This is the most rational of the three approaches, and the teams of three of the architects (B1, B4, C1) use this approach. In contrast to restricted evaluation based on a limited number of criteria, this approach allows the decision-maker to analyze all the pros and cons of the selected choice. The level of rationality achieved by this approach depends on humans' cognitive limitations, information availability, and other factors that affect the analysis.

- The first satisfactory choice: The remaining architects (A1, B3) said their teams use this approach, which involves the decision-makers taking the first available choice and evaluating it against the given context. If it is not rejected, they settle on that choice without looking for other suitable choices to compare it to.

Even though various metrics and measurements can be used to support decision-making, only two of the architects have been using any sort of measurements to assist decision-making. One of them uses information such as the code complexity and traceability matrix; another uses the metrics from a static code analysis tool for decision-making. No specific decision support tool was used during the process. However, whiteboards, Microsoft PowerPoint, and other drawing tools are used to facilitate communication.

Despite the fact that the majority of the teams employ some form of rational approach for decision-making, when it comes to the architects' own-decision making, only one of them claims to make decisions based on a methodological approach. As shown in Table IV, experience is the main source of support for decision-making, and it is closely followed by intuition. Prototyping possible solutions, using external experts and reusing available solutions are the other factors that decisions are based on. Both experience and intuition are personal qualities that are acquired by each individual over time. Prototyping is mostly used for evaluating and validating a selected solution.

TABLE IV. FACTORS AFFECTING ARCHITECT'S OWN DECISION-MAKING

| Based on / ID | A1 | A2 | A3 | A4 | B1 | B2 | B3 | B4 | C1 | C2 |
|---|---|---|---|---|---|---|---|---|---|---|
| Experience | X | X | X | X | X | X |  | X | X | X |
| External experts |  |  |  |  |  | X | X |  |  |  |
| Intuition | X | X |  | X | X | X | X | X |  |  |
| Prototyping |  |  | X | X |  |  | X |  | X |  |
| Methodology |  |  |  |  |  |  |  |  | X | X |
| Reusing | X |  | X |  |  |  |  |  |  |  |

Several issues that affect architecture decision-making were also mentioned during the interviews. One of the most frequent issues mentioned by the architects was maintaining the balance between time, cost, and quality. The architects are usually forced to make quick decisions due to time pressure, so they are unable to spend enough time searching for a better solution. Sometimes, due to the time and monetary constraints, they have to settle for easy-to-implement architecture instead of long-term benefits. It is usually hard to convince the customers and the higher management of the long-term benefits. Ambiguous and constantly changing requirements also hinder the decision-making process. Most of the architects manage to overcome this issue by staying in constant contact with customers.

Maintainability, reliability, testability, performance, and security are highlighted as the quality attributes most frequently considered during architecture decision-making. However, the majority of the architects claim to use those quality attributes as guidelines rather than using them as key decision factors. Even though some of the quality attributes are validated during testing, none of them conducts any extensive architecture evaluation to make sure that the designed architecture fulfills the quality requirements.

### C. Reasons for Using Current Decision-Making Approach (RQ2)

The majority of the software practitioners were happy with their current decision-making process even though they have recognized they need some improvements. None of them were dissatisfied with the current way of doing things, but three of them claim to be neutral about the current approach. According to the interviewees, the following are the main reasons for using their current way of decision-making.

- Lightweight process (A1, B4)
- Easy to reach consensus (B1, C1)
- Freedom to do things differently (A2, B2)
- Higher involvement and motivation among team members (A1, C2)
- Less documentation (A1)
- Robust process (A2)
- Responsiveness to clients (A4)
- Flexibility (A3)
- Faster process (B3)

Even though there are no widely accepted reasons for using the current architecture decision-making approach, there is a similarity among most of the reasons mentioned above. Based on that, it is possible to suggest that the majority of these architects consider their current approach either light, fast, or flexible, and they would like to hold on to the current approach. These reasons should be taken into account when developing architecture decision support for this context.

*D. Challenges Associated with Current Decision-Making Approach (RQ3)*

On the other hand, the interviewees have also recognized several challenges associated with their current decision-making approaches.

- Possibility of missing out on a better solution (A1, A2, A3, B1)
- Difficulty in revisiting the design rationale (A1, A2, C2)
- Problems related to integration of new members (A4, B4, C2)
- Improper documentation (B3, C1, C2)
- Issues with customer communication (A2)
- Knowledge gap between the engineers (A4)
- Difficulty in finding the necessary resources (B2)
- Lack of proper tools (C1)

The challenges associated with the current decision-making approach highlight the downside of using an informal approach. The possibility of missing out on a better solution, a challenge mentioned by four participants, is clearly the result of lack of a systematic process. The majority of the remaining challenges appear to be related to the lack of proper knowledge management.

*E. Possible Improvement Aspects (RQ4)*

The software architects were asked to state their wishes for a solution that improves architecture decision-making. The following are the improvement ideas most frequently mentioned during the interviews.

- Lightweight technique or tool to guide them (A1, A2, A3, C1, C2)
- Improved documentation (A2, B2)
- Efficient information sharing (A4, B1, C1)
- Keeping track of design decisions and rationale (B1, B2)
- Making the decision-making more agile (B2, B3)

Developing a lightweight technique or tool to guide them during the architecture decision-making process was one of the main requests. Since the architects indicated their resistance to using process-heavy tools throughout the study, this appears to be a natural choice. Even though they understand the shortcomings of the current process, they are not willing to replace it with a time-consuming process because most of the projects are already facing time management issues. The rest of the improvement requests can be considered as characteristics of the expected solution, and all of them are related to architecture knowledge management. Hence, we further analyzed the results to derive the architecture knowledge management practices at the team level as well as at the organizational level.

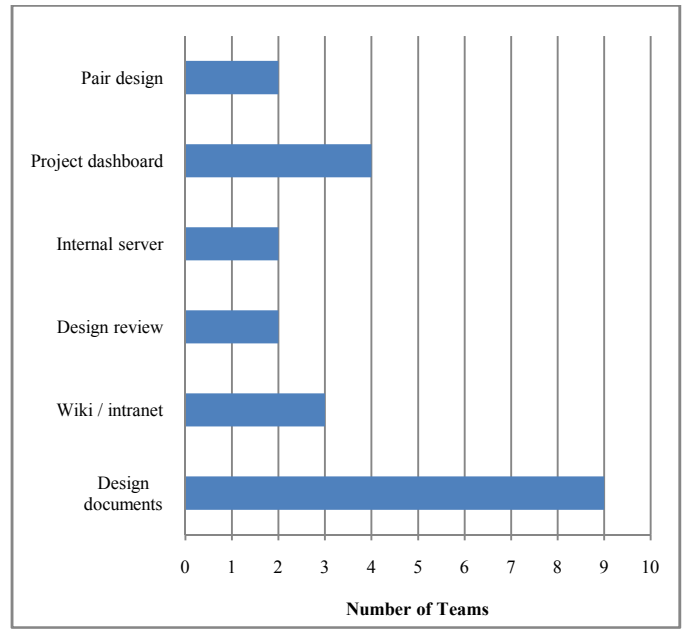

Fig. 2. Architecture Knowledge Management Techniques

The need for knowledge management frequently surfaced during the interviews. While some of the activities were specifically recognized as knowledge management activities by the interviewees, others were derived from different parts of the discussion since their role as knowledge management activities was not obvious to the architects. Fig. 2 shows the identified knowledge management activities and the number of teams that engage in these activities.

Recording information in the design documents is the most frequently used knowledge management technique. However, most of the participants admitted that they have several issues related to the design documents regarding their quality and maintenance. While most design documents follow a general template and update procedure, design justification documents and evolution analysis documents are two notable exceptions. They are company-specific documents that are intended to capture the design rationale for each design decision made. Though the architects see the value of recording the design rationale, the use of these documents is restricted to limited areas of the corresponding companies. While most knowledge management activities are related to documentation, pair design and design review are two practical tasks that help in managing tacit knowledge. Several other informal knowledge management activities are used all the teams. Swapping tasks, customer interactions, meetings, brainstorming sessions, and informal discussions are some of the activities that contribute to knowledge management at different levels.

V. DISCUSSION

This study highlights several contextual characteristics that make software architecture decision-making a challenging task. Klein et al. [20] has presented several situation characteristics: ill-structured tasks, ambiguity and missing data, shifting and competing goals, dynamic conditions, action-feedback loops, time stress, high stakes, multiple players, and organizational goals and norms that make it difficult for decision-makers to analyze all available options and take the optimal cause of action in complex real-world scenarios. All the characteristics mentioned above were identified in the studied context. Unclear and unstructured requirements; constant change requests; rapid technological changes; conflicting stakeholder concerns; balancing between time, cost, and quality; and the challenges related to the organizational practices are some of the corresponding issues that came up during the study. Those issues can also be mapped onto the factors that are recognized as the characteristics that make architecture decision-making a difficult task [11].

Despite the availability of a considerable number of architecture decision-making techniques, the study showed that none of the software teams use a systematic approach to make architecture decisions in real-life contexts. However, they use informal but structured approaches that share similar characteristics with some of the systematic techniques. Among the three recognized approaches followed, selecting a choice that fulfills pre-defined criteria is similar to the Quality-Drive Decision Support Method [41], selecting a choice after analyzing pros and cons has the same characteristics as CBAM [40], and selecting the first satisfactory choice follows the same approach as the Recognition Primed Decision Model (RPD) [20]. The findings of this study are in line with the findings of the recent studies conducted by van Heesh et al. [12] and Anvaari et al. [13], in which researchers came to the conclusion that the majority of teams use their own customized decision-making approaches rather than using a systematic approach from the literature.

Software architects' own decision-making during architecting appears to be heavily based on personal qualities rather than external resources. Intuition and experience have been recognized as key drivers in software architecture decision-making [41], and our study clearly confirms their prominent role during the process. However, the decision-makers are prone to making errors when make decisions under uncertainty due to heuristics and biases [19]. Hence, using both tacit and explicit knowledge together would help them to make better decisions. The lack of systematic use of explicit knowledge is quite evident, and the increased use of explicit knowledge would help improve decision-making. Moreover, the identified factors that affect the architects' decision-making can be categorized into three different areas that are recognized as the main drivers of software architecture design—reuse, method, and intuition [7]. Experience and intuition can be placed in one category while using external experts and methodology is in the other. The third category would consist of reuse. Prototyping can be placed in any of the categories as it has the qualities of all of them.

Even though the architects identified several challenges related to their current decision-making approaches, most of them are satisfied with their current practice because it is lightweight and flexible. As discussed previously, the context in which architecture decisions are made is very demanding, and the architects prefer to compromise on some of the qualities in order to have less overhead. However, it is possible that the reasons that are considered advantages negatively affect the architecture as well as the decision-making process itself. For example, even though less documentation is considered an advantage, it hinders recording design decisions and decision rationale. While it makes the current decision-making process faster, the inability to revisit design decisions and decision rationale can have negative consequences in later stages of the development process [46].

Though one of the main challenges associated with the current decision-making approach is a result of lack of a systematic process, the majority of the remaining challenges can be linked to improper knowledge management. The possible improvement aspects yield a similar result since, except for the request for a lightweight tool/technique to guide the process, the rest of the improvement aspects are also related to architecture knowledge management.

In analyzing the results it is clear that architecture decision-making can be positively influenced by improving architecture knowledge management because it would address several identified challenges as well as fulfilling recognized improvement aspects. Improving documentation is one of the activities that can significantly contribute to improving architecture knowledge management. Since the practitioners are generally familiar with documentation, implementing good documentation practices is quite straightforward. Nevertheless, maintaining better documentation practices for a longer run is a challenging task [46]. As we noticed during the study, even though the practitioners are aware of the general benefits of other architecture knowledge management activities, there is no systematic knowledge management process in any of the companies. To give a better understanding of current architecture knowledge management activities followed in these companies, they are mapped onto the SECI model [32] as shown in Fig. 3.

|  | Tacit Knowledge To | Explicit Knowledge |
|---|---|---|
| From Tacit Knowledge | **Socialization**<br>• Pair design<br>• Design review<br>• Swapping tasks<br>• Customer interactions<br>• Informal discussions | **Externalization**<br>• Meetings<br>• Brainstorming<br>• Retrospective |
| Explicit Knowledge | **Internalization**<br>• Prototyping | **Combination**<br>• Design documents<br>• Wiki / intranet<br>• Internal server |

Fig. 3. Architecture KM activities mapped onto the SECI model

Each identified architecture knowledge management activity was placed with the most relevant type of knowledge conversion presented in the SECI model. As shown in the model, several socialization activities that transfer tacit knowledge among stakeholders take place in the companies. Even though those activities are not purposefully implemented for knowledge conversion, they help to share knowledge. Documenting architecture knowledge in design documentation and in wikis contributes to combination, in which scattered architecture knowledge is combined and made available to the stakeholders. Externalization that converts tacit knowledge into explicit knowledge is supported by retrospective meetings and brainstorming sessions. Finally, prototyping is recognized as an internalization activity since it helps the person who builds the prototype to acquire tacit knowledge. Even though several other activities can aid knowledge conversion, these are the activities that were recognized as the main contributors. Improving existing architecture knowledge management activities and introducing new architecture knowledge management activities will lead to better software architecture decisions.

## VI. Threats To Validity

The way the interview questions were constructed and presented could cause a threat to the validity of the study as it could affect the interviewees' answers. This was taken into consideration during the case study design and several steps were taken to minimize the effect. The interview questions were designed to be mostly open-ended and they were used as guidance rather than as a questionnaire. The interview guide also went through several rounds of improvements based on the feedback from the pilot study participant as well as the senior researchers who evaluated it. Since all the case companies are SMEs and have similar characteristics, generalizing the results to be applied at a different context will be difficult. A replication of the case study at a large-scale enterprise has been planned as a means to increase the generalizability.

## VII. Conclusion

This study provides several insights about the software architecture decision-making practices in three SMEs. It reveals that the software architects in the given context don't follow any systematic software architecture decision-making technique to make architecture decisions. Instead they follow informal but structured approaches for team level architecture decision-making. Interestingly, the approaches used by the architects closely resemble some of the existing systematic decision-making techniques despite being much lighter than their process-heavy counterparts. This indicates that the practitioners might be willing to adopt existing solutions if they were made lightweight while maintaining the underlying decision-making process. The individual level decision-making was mostly based on personal characteristics such as intuition and experience. We also identified several factors, including time, money and organizational practices that make it difficult for decision-makers to conduct extensive analysis before making a decision. Architecture knowledge management is identified as one of the main aspects that should be improved in order to have better architecture decisions. Our next goal is to test the validity of the findings of this study in a different context. We are planning to conduct a case study in a large enterprise where there are larger software projects, a complex organizational structure, and a large number of stakeholders that make the decision-making process much more complicated. The findings will be compared with the results of this study.

### Acknowledgment

We would like to thank all the interviewees and the case companies for their cooperation throughout the study. This work is funded by ITEA2 and Tekes - the Finnish Funding Agency for Innovation, via the MERgE project, which we gratefully acknowledge.